\documentclass{article}

\usepackage{arxiv}

\usepackage[utf8]{inputenc} 
\usepackage[T1]{fontenc}    
\usepackage{hyperref}       
\usepackage{url}            
\usepackage{booktabs}       
\usepackage{amsfonts}       
\usepackage{nicefrac}       
\usepackage{microtype}      
\usepackage{lipsum}
\usepackage{graphicx}
\graphicspath{ {./images/} }
\usepackage{amssymb}
\usepackage{amsmath}
\usepackage{multirow}
\usepackage{setspace}
\usepackage{algorithm}
\usepackage{algpseudocode}
\usepackage{threeparttable}
\title{100Mbps Reconciliation for Quantum Key Distribution Using a Single Graphics Processing Unit}

\author{
 Yu Guo$^{*}$ \\
  State Key Laboratory for Novel Software Technology, \\
  Nanjing University, Nanjing,\\ 
  210046, China\\
  \texttt{mg1833101@smail.nju.edu.cn} \\
   \And
 Chaohui Gao$^{*}$\\
  Nanjing University, Nanjing,\\ 
  210046, China\\
  * Authors contributed equally to this work \\
  \texttt{m13578756629@163.com} \\
  \And
 Dong Jiang \\
  State Key Laboratory for Novel Software Technology, \\
  Nanjing University, Nanjing,\\ 
  210046, China\\
  \texttt{jiangd@nju.edu.cn} \\
   \And
  Lijun Chen \\
  State Key Laboratory for Novel Software Technology, \\
  Nanjing University, Nanjing,\\ 
  210046, China\\
  \texttt{chenlj@nju.edu.cn} \\
}

\begin{document}
\maketitle
\begin{abstract}
An efficient error reconciliation scheme is important for post-processing of quantum key distribution (QKD). Recently, a multi-matrix low-density parity-check codes based reconciliation algorithm which can provide remarkable perspectives for high efficiency information reconciliation was proposed. This paper concerns the improvement of reconciliation performance. Multi-matrix algorithm is implemented and optimized on the graphics processing unit (GPU) to obtain high reconciliation throughput. Experimental results indicate that GPU-based algorithm can highly improve reconciliation throughput to an average \textbf{85.67} Mbps and a maximum \textbf{102.084} Mbps with typical code rate and efficiency. This is the best performance of reconciliation on GPU platform to our knowledge.
\end{abstract}


\section{Introduction}
Quantum key distribution (QKD) allows two legitimate parts, Alice and Bob, to share unconditional secure keys through quantum channel and classical channel \cite{bennett2014quantum,gisin2002quantum}. Based on the physical principles and theories, QKD guarantees that keys are safe against eavesdropper Eve. Normally, QKD process is divided into two phases: quantum phase and post-processing classical phase. In the first phase, Alice and Bob obtain the raw key through the quantum channel respectively.

In the second phase, in order to ensure that Alice and Bob share the same key, Bob’s raw key needs to be corrected, and the post-processing is introduced. Post-processing includes four stages: base sifting, error estimation, reconciliation \cite{chung2001analysis,kou2001low} and privacy amplification \cite{bennett1988privacy,bennett1995generalized}. Before reconciliation, Alice and Bob will get sifted key respectively. Then in reconciliation stage, Bob corrects the errors using the reconciliation algorithm to assure the consistency between their sifted key. In this stage, Alice and Bob should leak as less information as possible in this stage. The performance of reconciliation is the bottleneck of the QKD system. So the scope of this paper lies in reconciliation scheme establishing with appropriate algorithms and platforms.
	
Most researches adopt belief propagation (\textbf{BP}) as reconciliation algorithm \cite{chung2001analysis,kou2001low}. Traditional \textbf{BP} algorithm uses one low-density parity-check (LDPC) code to correct the errors \cite{kiktenko2018error}. And \textbf{BP} is implemented on CPU, GPU and FPGA to breakthrough performance bottleneck \cite{dixon2014high,wang2018high,yuan201810,mao2019high}. The ultimate goal of these studies is to improve the reconciliation performance.
	
Recently, a highly efficient multi-matrix algorithm was proposed by Gao et.al in Ref. \cite{gao2019multi}, in which they use two or more matrices as check matrix. This algorithm was proved to be efficient and safe, it can reduce the number of iterations, increase the success rate of reconciliation and reduce the error rate after reconciliation \cite{gao2019multi}. This algorithm generates multiple syndromes to transmit and update data, and these processes involves a host of repetitive, simple operations. Furthermore, compared to other platforms, GPU is suitable for processing data-intensive computing tasks. So in order to take advantage of their similarity, we decide to realize this novel reconciliation algorithm on GPU to obtain better reconciliation performance.
	
In this paper, we design and realize a novel reconciliation scheme using multi-matrix algorithm on GPU platform, called multi-matrix scheme. The primary purpose of this scheme is to break through the bottleneck of reconciliation performance. 
Experimental results demonstrate that multi-matrix scheme has higher throughput, success rate and lower number of iterations. After further optimization, this scheme can be extended to any key length and GPU platform. Moreover, this scheme can achieve better performance with typical code rate and efficiency. It is promising that our multi-matrix scheme is applicable in real-time QKD system \cite{yuan201810,korzh2015provably}. 
	
The rest of this paper is organized as follows: in section 2, we give a related review of the multi-matrix algorithm and some parameter definitions. In section 3, we describe how we adopt the multi-matrix algorithm on GPU and the optimization of algorithm in detail. In section 4, experimental results are listed to show the high performance of multi-matrix algorithm. In section 5, we draw a conclusion and summary of our work.
	
\section{Preliminaries}
Multi-matrix \textbf{BP}, called \textbf{MBP}, use \emph{u}($\emph{u}>1$) matrices generated by \textbf{PEG} algorithm \cite{hu2005regular,hu2001progressive} to correct errors simultaneously. The dimensions of these matrices are $m*n$, where $n$ is the length of the key and the number of variable nodes, and $m$ the number of check nodes.

Before reconciliation, Alice and Bob get their sifted key $\textbf{$\textbf{x}^{T}$}=[x_{1},x_{2},...,x_{\textbf{n}}]$ and 
$\textbf{$\textbf{y}^{T}$}=[y_{1},y_{2},...,y_{\textbf{n}}]$ ($x_{i},y_{i}\in \{0,1\}$) respectively. Alice and Bob share \emph{u} matrices $\emph{$H_{1}$}$,$\emph{$H_{2}$}$,...,$\emph{$H_{\emph{u}}$}$. Alice calculates $\emph{u}$ syndromes ${(\textbf{z}^{l})}^{T}$ according to the formula(1), and sends them to Bob through the classical channel.
	\begin{equation}
		{(\textbf{z}^{l})}^{T}=[z_{1},...,z_{\textbf{m}}]=\emph{H}_{l} \cdot \textbf{x} (mod 2),l\in \{1,2,..,u\},z_{j}\in \{0,1\}
		\label{eq:refname1}
	\end{equation}
	
Then Bob initializes the prior probabilities $\emph{P}_{i}^{b}$($b \in \{0,1\}$), and then calculate the log likelihood ratios $\emph{L}^{l}_{P_{i}}$ and variable-to-check (V2C) information $\emph{L}^{l}_{C_{j}\to V_{i}}$ for all \emph{u} matrices respectively. After this, Bob updates and propagates check-to-variable (C2V) information. Finally, Bob goes through all variable nodes to get the value $\emph{L}^{l}_{V_{i}}$ of soft-decision according to formula(2), where $N_{l}(V_{i})$ represents the set of adjacent variable nodes of check nodes ${V_{i}}$ in $l^{th}$ matrix. 
	\begin{equation}
		\emph{L}_{V_{i}}=\emph{L}_{P_{i}}+\sum_{l=1}^{u} \sum_{C_{j}\in N_{l}(V_{i})} \emph{L}^{l}_{C_{j}\to V_{i}}
		\label{eq:refname3}
	\end{equation}
And $\textbf{$\textbf{y}^{T}$}$ will be corrected by decoding decisions. Once ${\textbf{z}^{l}}= \emph{H}_{l} \cdot \textbf{y}$ (mod 2), reconciliation is considered to be successful and finished. Otherwise, the reconciliation continues until the number of iterations overtake the limit, which signifies a failing reconciliation. Detailed reconciliation process and algorithm can be found in Ref.\cite{gao2019multi}.

The reconciliation efficiency \emph{f} is an important parameter\cite{elkouss2009efficient}, which shows the ratio of the actual amount of information Bob obtains to the theoretical minimum amount of information Bob needs for correcting all errors. In real applications, \emph{f} is calculated according to formula (3), which is same as the single-matrix algorithm,
	
	\begin{equation}
		\emph{f}=\frac{{m}}{{n}h(\emph{e})}>1,
		\label{eq:refname3}
	\end{equation}
where \emph{e} is the error rate of key and \emph{h(e)} is the Shannon binary entropy:
	
	\begin{equation}
		h(\emph{e})=-\emph{e}log_{2}\emph{e}-(1-\emph{e})log_{2}(1-\emph{e})
		\label{eq:refname3}
	\end{equation}
In the real QKD system, the value of \emph{f} stages around 1.1 to 1.4. The lower \emph{f}, the less information needs to be shrink in privacy amplification stage, and vice versa.

\section{Scheme implementation and optimization}

	\noindent\textbf{A.Implementation}\\
	\textbf{MBP} is shown in \textbf{Algorithm 1}. 
There are $n$ nodes needed to be updated in every matrix in \textbf{MBP}, and the process of updating is similar. Meanwhile, GPU can call thousands of threads to do some simple, repetitive calculations in parallel\cite{kirk2007nvidia}, so nodes can be updated simultaneously.
	
	\begin{algorithm}
		\setstretch{1.2} 
		\caption{Multi-matrix reconciliation algorithm}\label{alg:euclid}
		\begin{algorithmic}[1]
			\State Initialize $L^{l}_{V_{i}\to C_{j}}=L^{l}_{P_{i}}$
			\For{every parity-check matrix $H_{l}$}
			\For{j=1 to m}
			\For{every $V_{i}^{k} \in $neighborhood of $C_{j}^{k}$ }
			\State Generate and propagate  $L^{l}_{C_{j}\to V_{i}}$
			\EndFor
			\EndFor
			\For{i=1 to n}
			\For{every $C_{j}^{k}\in $ neighborhood of $V_{i}^{k}$}
			\State Generate and propagate $L^{l}_{V_{i}\to C_{j}}$
			\EndFor
			\EndFor
			\EndFor
			\State Make decoding decisions
			\If{stopping rule is not satisfied}
			\State{Go back to line2}
			\EndIf
		\end{algorithmic}
	\end{algorithm}
In our scheme,  CPU is responsible for obtaining the matrices and key information and calling GPU. Then GPU allocate threads to start updating node information. We decide to use one thread to update one information of node. However, GPU cannot call all threads at the same time. Instead, GPU treats Warp as its basic execution unit. There are 32 threads in a Wrap, the basic execution unit of GPU. All threads in the same Wrap execute simultaneously. And GPU can call up to 80 Wraps to work at the same time\cite{kirk2007nvidia}, and it will achieve the best performance when the number of called threads is a multiple of 32, i.e. $n$ mod 32 = 0. Our sifted key length is $2^{20}$, leading to the huge and inflexible matrix. If all matrices and keys are stored in GPU, storage and computing resources will be greatly wasted.
	
So we split the sifted key into $k$ short keys of length $n$, i.e. $k*n$=$2^{20}$. And one thread updates information of one node, in other word, $n$ nodes need at least $n$ threads to be updated. In this way, the size of matrices changes from $m*2^{20}$ to $m*n$, which greatly reduces the use of storage resources.  The process reconciliation on GPU sees in \textbf{Fig. 1}. \\
	
	\begin{figure}[htbp]
		\centering
		\fbox{\includegraphics[width=10cm,height=7cm]{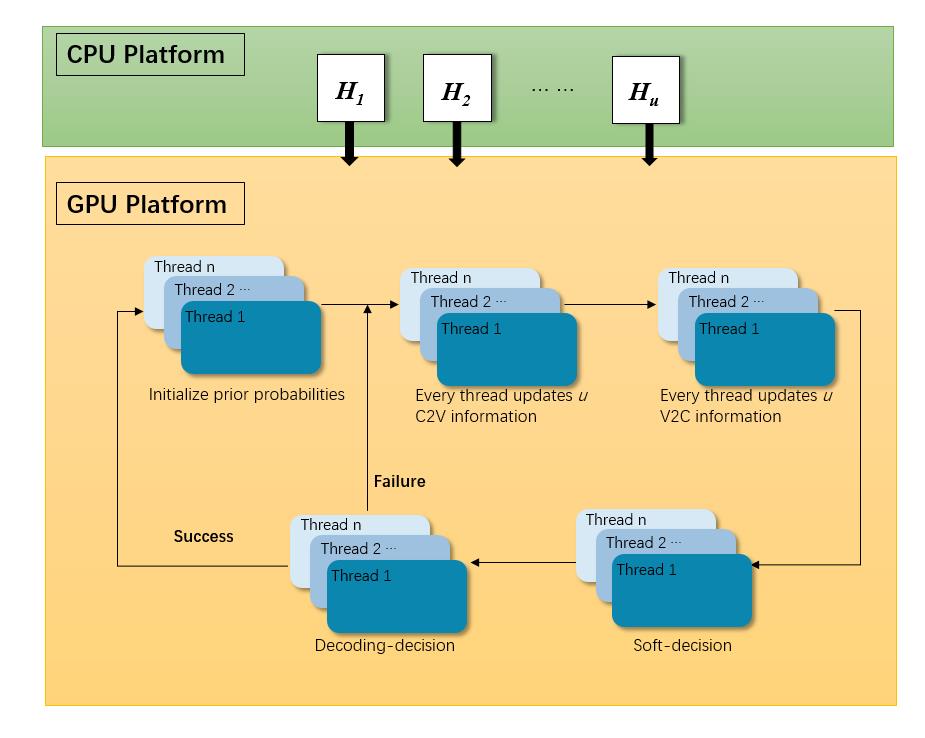}}
		\caption{Process of our multi-matrix scheme on GPU best in color.}
		\label{fig:false-color}
	\end{figure}

	\noindent\textbf{B.GPU optimization}\\
In this work, we optimize GPU and \textbf{MBP} from the following three aspects to obtain the best performance: thread optimization, branch reduction, and memory optimization.

\textbf{Thread optimization}. In the implementation process, the sifted key is splited into short keys, which not only makes multi-matrix scheme suitable for various key length, but also enables GPU to allocate threads reasonably. GPU allows users to divide threads into multidimensional forms logically, and different division modes affect the performance of GPU\cite{kirk2007nvidia}. So we need to find the best division mode according to parameter $n$. We schedule the thread blocks in two dimensions, denoted as \emph{Block($B_{x}$,$B_{y}$)}, and we schedule threads in two dimensions in every thread block, denoted as \emph{Thread($T_{x}$,$T_{y}$)}. So in \textbf{Fig. 1}, all shown threads are organized two-dimensional forms in two-dimensional thread blocks logically. We test the throughput of different split methods, as shown in \textbf{Table 1}. As a result, we decide to use the last method, where $n$=$2^{16}$, $k$= 16, \emph{Block($B_{x}$,$B_{y}$)}=(16,8), \emph{Thread($T_{x}$,$T_{y}$)}=(32,16) to obtain the best performance.

	\begin{table}[htbp]
		\setstretch{1.5}
		\centering
		\caption{\bf Thread scheduling and throughput}
		\begin{tabular}{cccc}
			\hline
			$n$ & $Block$ & $Thread$ &  Throughput(Mbps) \\
			\hline
			$10^{4}$     & $(16,2)$        & $(10,32)$      & $51.266$\\
			$2^{14}$     & $(16,16)$       & $(8,8)$        & $55.241$\\
			$2^{15}$     & $(16,16)$       & $(16,8)$       & $72.959$\\
			$2^{16}$     & $(2,2^{15})$    & $(1,1)$        & $61.326$\\
			$2^{16}$     & $(16,8)$        & $(32,16)$      & $102.084$\\
			\hline
		\end{tabular}
		\label{tab:shape-functions}
	\end{table}
	
\textbf{Branch reduction}. Each thread in the same Wrap executes the same instructions. Threads with different branches in the same Warp, leading to the different instructions, will wait for each other. Those statements with branches, such as \textbf{if} and \textbf{for} statements, will reduce GPU parallelism and reconciliation throughput\cite{kirk2007nvidia}. There are numerous branches in \textbf{MBP} and most of them can be trimmed.  In our scheme, these branches are reduced as much as possible, and the \textbf{MBP} is optimized to accommodate the GPU architecture. For example, in \textbf{MBP}, some judge statements are replaced by some general expressions. We initialize the prior probabilities $\emph{P}_{i}^{b}$($b \in \{0,1\}$) according to formula(5). 
	
	\begin{equation}
	\setstretch{1}
	\left\{
	\begin{aligned}
	P_{i}^{1}=-(y_{i}-{e})& \\
	P_{i}^{0}=1+(y_{i}-{e})
	\end{aligned}
	\right.
	\end{equation}
And we propagate check-to-variable(C2V) information $\emph{L}_{C_{j}\to V_{i}}$ according to formula (6), where $v_{i^{'}}\in N(C_{j})\backslash i$ represents that $v_{i}$ is not included in the set. Using similar approaches, over 80$\%$ branches are trimmed, which improves reconciliation performance by over 10$\%$.
	
	\begin{equation}
		\setstretch{1}
		\emph{L}_{C_{j}\to V_{i}}=(-1)^{z_{j}} \cdot  2tanh^{-1} 
		\left(\prod_{v_{i^{'}}\in N(C_{j})\backslash i}
		tanh\left(\frac{1}{2}\emph{L}_{v_{i^{'}}\to{C_{j}}}    \right)    \right)
	\end{equation}
	
\textbf{Memory optimization.} All matrix node information is initially stored in CPU. GPU interacts with CPU to read and write data. Usually, there will be an interaction between GPU and CPU before the next iteration begins. Through our tests, the interaction time exceeds the calculated time in one iteration. So before reconciliation, all nodes of matrices and keys are put in the global memory in GPU, and all intermediate results generated in one iteration are put in share memory and cache. During reconciliation, threads send signals to each other to synchronize data. In this way, the interaction between GPU and CPU only occur at the beginning and the end of reconciliation, leading to the improvement of GPU computing resource utilization. Also, we use coalesced global memory, which is also applied in \cite{wang2018high}, to hide the latency of global memory. \\
	
\section{Experimental results}
	
\textbf{Optimization results}. After the optimization in section 3, multi-matrix scheme implemented on the GPU achieves best reconciliation performance. Our scheme makes full use of available threads and schedules them appropriately. Also, this scheme reduces the use of memory and average iteration times. Every optimization reduce reconciliation time leading to the higher throughput, as shown in \textbf{Table. 2}. In this table, we compare several parameters before and after optimization. And the optimization results of each aspect are also listed in the \textbf{Table.2}.
	
\begin{table}[htbp]
		\setstretch{1.5}
		\centering
		\caption{\bf Performance Comparison}
		\begin{threeparttable}
			\begin{tabular}{c|c|c}
				\hline
				Parameter& Before $\textbf{OP}^{1}$ &  After $\textbf{OP}^{1}$ \\
				\hline
				Least Threads                  & $u*n*k$                 & $n$   \\
				Space occupation              & $u*n*k$                  & $u*n$                   \\
				Average iterations             & $5.16$             & $3.71$           \\
				Average iteration time(ms)     & $293.6$            & $164.4$         \\
				\hline
				Factor&\multicolumn{2}{c}{Throughput(Mbps)}\\
				\hline
				Thread            	 &\multirow{4}*{48.672}  		 &$69.260$		\\
				Branch				&~								 &$53.973$		\\
				Memory 					&~   				 		   	&$70.426$		\\
				Best Results       	&~		           				& $102.084$    \\
				
				\hline
			\end{tabular}
			\label{tab:shape-functions}
			\begin{tablenotes}
				\footnotesize
				\item[1] $\textbf{OP}$: optimization
			\end{tablenotes}
		\end{threeparttable}
\end{table}

\textbf{Improvement of multi-matrix}. Multi-matrix algorithm has a significant improvement of reconciliation, i.e. faster reconciliation speed, higher success rate. In \textbf{Fig. 2}, we compare the throughput and success rate of the different schemes. Experimental results that multi-matrix scheme has higher throughput and success rate than single-matrix scheme with the increase of \emph{e}.  And after our optimization for algorithm and GPU, the performance of reconciliation is further improved. 
	\begin{figure}[htbp]
		\centering
		\fbox{\includegraphics[width=10cm]{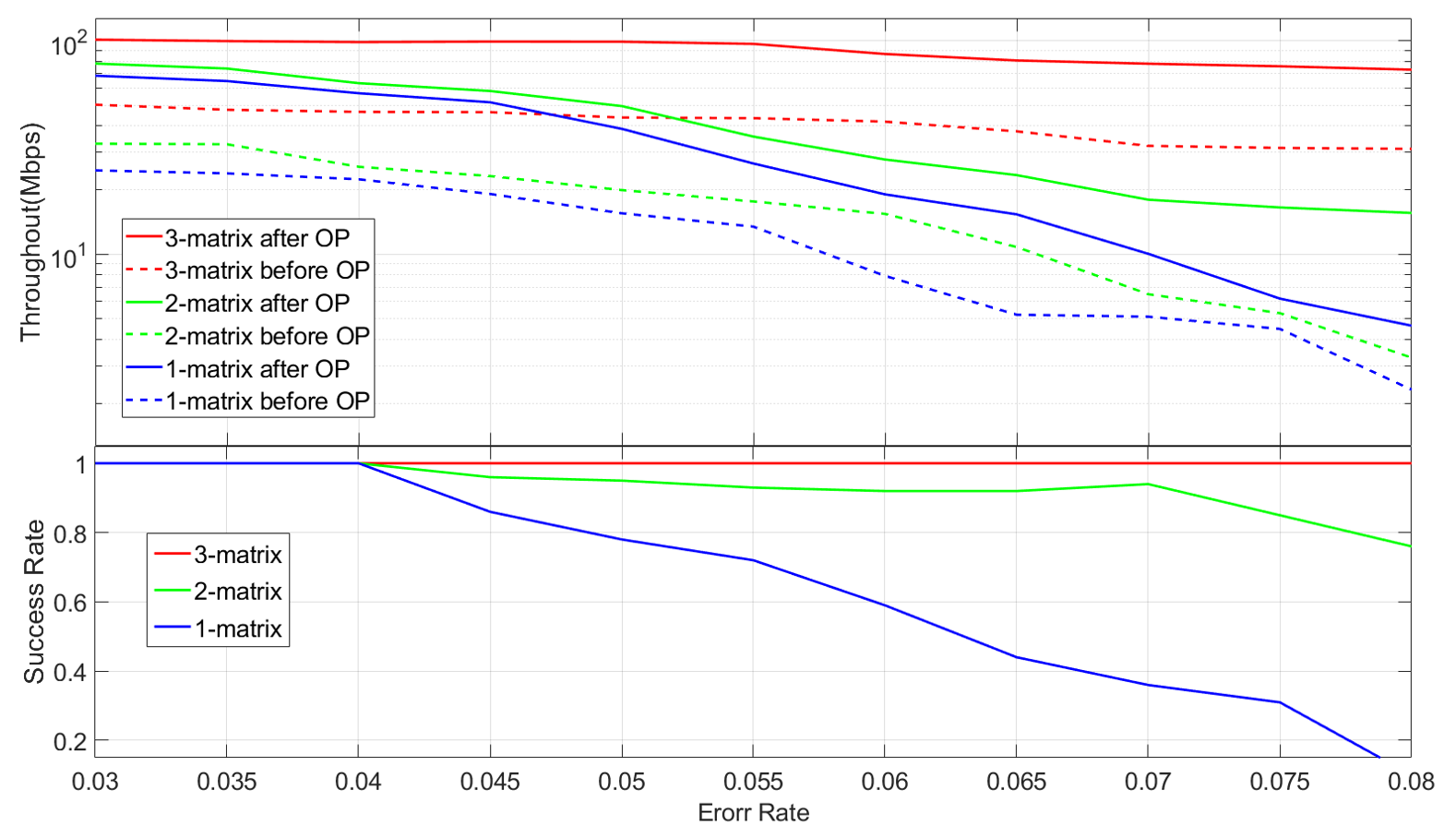}}
		\caption{Throughput different by number of matrices and whether optimized, where error rate \emph{e} is from 0.03 to 0.1, and code rate \textbf{R}=0.5.}
		\label{fig:false-color}
	\end{figure}
	
\textbf{Fig. 2} displays the throughput using single-matrix scheme, 2-matrix scheme and 3-matrix scheme before and after optimization respectively. The figure clearly demonstrates that throughput increases as the number of matrices increases and the overall reconciliation performance is greatly improved after optimization.  Through our tests, the ability of error correction of multi-matrix algorithm approaches saturation as the number of matrices increases. So considering computing resources, algorithm performance and other factors, we conclude that the reconciliation system achieves the best performance when the number of matrices is 3, i.e. \emph{u}=3.
	
\textbf{Usability}. Multi-matrix scheme can apply to various code rate $\textbf{R}$, which is defined as $\textbf{R}=1-\frac{\textbf{m}}{\textbf{n}}$. We choose 4 typical \textbf{R}, i.e. 0.5, 0.6, 0.7, 0.8, and test the throughput, as shown in \textbf{Fig. 3}. Optimized scheme is conducted on TitanV and GTX 1060, different models of NVIDIA GPU. The experimental results imply that multi-matrix scheme can get promising performance on different GPU platform. 
	
	\begin{figure}[htbp]
		\centering
		\fbox{\includegraphics[width=10cm,height=7cm]{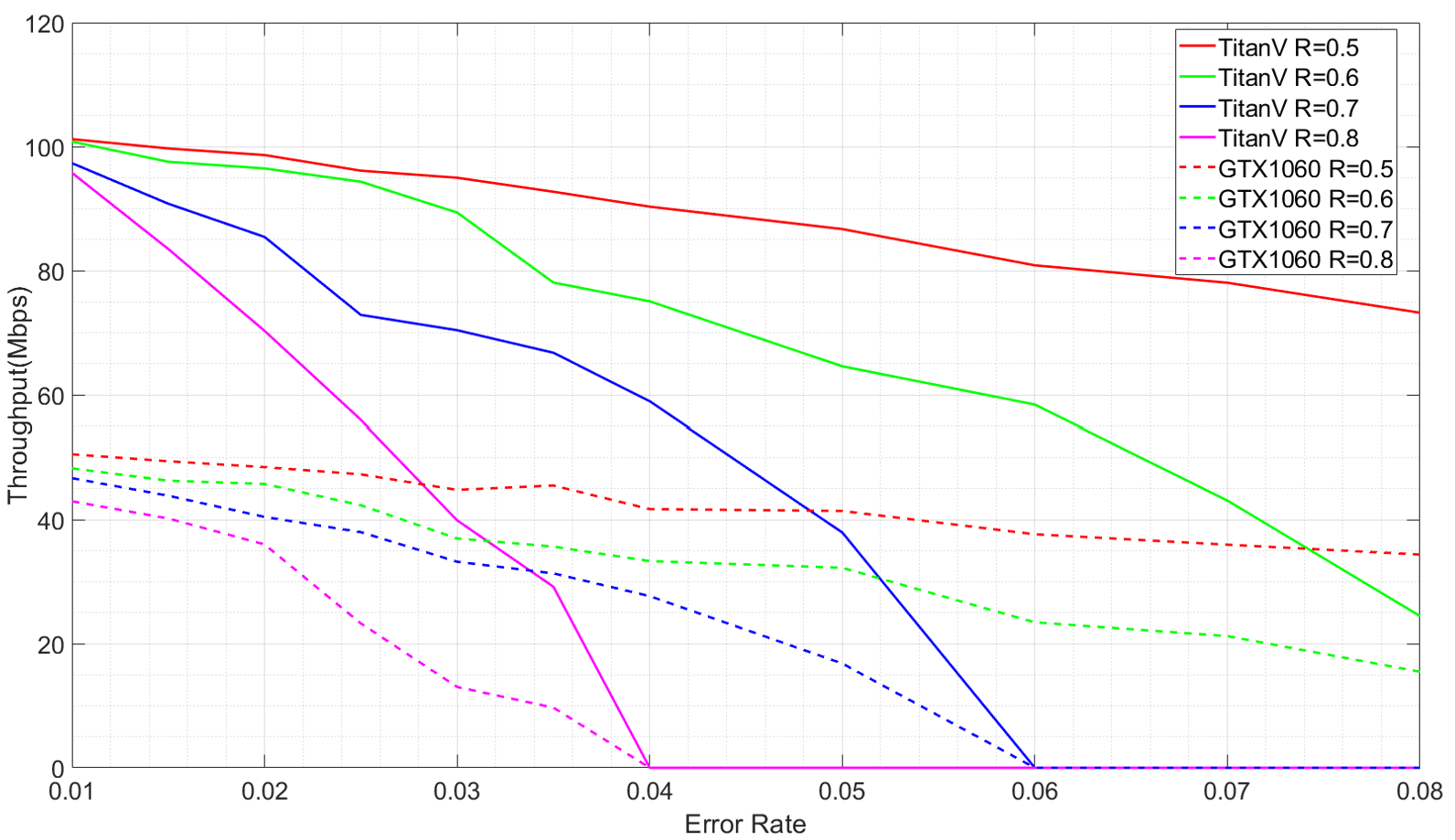}}
		\caption{Throughput with different code rate \textbf{R}.}
		\label{fig:false-color}
	\end{figure}
\textbf{Fig. 3} indicates that the amount of information leaked during the reconciliation is acceptable with high reconciliation throughput. What's more, multi-matrix scheme is easier to correct the errors with high \emph{e} than single-matrix scheme. So it can achieve favorable efficiency with high throughput, as is demonstrated in \textbf{Table 3}, which means less information needs to be shrunk in the privacy amplification stage.
	
	
	\begin{table}[htbp]
		\setstretch{1.5}
		\centering
		\caption{\bf Efficiency and Throughput}
		\begin{threeparttable}
			\begin{tabular}{cccc}
				\hline
				Efficiency & $\textbf{R}^{1}$&  $\textbf{T}^{2}$ &Average iterations \\
				\hline
				$1.1\leq f < 1.15 $     & 0.5              & $65.578$          & $6.65$          \\
				$1.15\leq f < 1.2 $       & 0.5              & $82.789$           & $5.00$              \\
				$1.2\leq f < 1.4$        & 0.5              & $95.720$         & $3.98$            \\
				$1.1 \leq f < 1.15 $      & 0.6              & $69.626$       & $6.60$ \\
				$1.1 \leq f < 1.15$       & 0.7              & $61.903$       & $6.67$ \\
				$1.1 \leq f <1.15$       & 0.8              & $64.820$        & $5.97$ \\
				\hline
			\end{tabular}
			\label{tab:shape-functions}
			\begin{tablenotes}
				\footnotesize
				\item[1] $\textbf{R}$: Code Rate.
				\item[2] $\textbf{T}$: Throughput (Mbps)
			\end{tablenotes}
		\end{threeparttable}
	\end{table}

\textbf{Comparison}. This experiment compares  throughput in different references, which all use single-matrix algorithm. In \textbf{Table 4}, we compare some typical researches on different platforms with our work. We keep reconciliation efficiency in the same range and compare the reconciliation throughput. Our multi-matrix scheme performs best to our knowledge.
	
	\begin{table}[htbp]
		\setstretch{1.5}
		\centering
		\caption{\bf  Throughput comparison}
		\begin{threeparttable}
			\begin{tabular}{cccc}
				\hline
				Ref.&   Platform &  $\emph{f}^{1}$ &$\textbf{T}^{2}$  \\
				\hline
				\cite{dixon2014high}     & GPU NVidia M2090           & $1.25$        &   $30.70$     \\
				\cite{dixon2014high}       & CPU Inter X5675                & $1.25$                 &   $9.00$         \\
				\cite{wang2018high}     & GPU TitanXp(CV-QKD)$^{3}$            & $0.93$                 &   $30.39$       \\
				\cite{yuan201810}      & FPGA Altera Stratix V           & $1.13\leq f < 1.2$      &   $55.00$        \\
				\cite{mao2019high}      & CPU i7-6700HQ                   & $1.108$                &  $57.60$        \\
				Our work         & GPU TitanV                      & $1.1\leq f < 1.2$      &   $70.13$ \\
				Our work         & GPU TitanV                      & $1.1\leq f < 1.3$      &   $85.67$ \\
				\hline
			\end{tabular}
			\label{tab:shape-functions}
			\begin{tablenotes}
				\footnotesize
				\item[1] $\emph{f}$: efficiency.
				\item[2] $\textbf{T}$: Throughput (Mbps).
				\item[3] Our works and others are based on DV-QKD.
			\end{tablenotes}
		\end{threeparttable}
	\end{table}

\section{Conclusion}
In this paper, a novel multi-matrix reconciliation algorithm is implemented and optimized on the GPU platform. Optimized multi-matrix scheme achieve better performance compared to single-matrix scheme, especially as the number of matrices increases. Under the premise of ensuring the reconciliation efficiency, we conduct our experiments using matrices of multiple code rate. Experimental results show that our multi-matrix scheme is the best scheme based on GPU to our knowledge. In addition, according to Ref. \cite{yuan201810}, our scheme is suitable for the real-world QKD system. After analysis, our scheme can improve 27\% to 40\% key rate using the QKD system in Ref. \cite{yuan201810}.\\
	
\noindent \begin{LARGE} \textbf{Funding}\end{LARGE}\\
	This research is financially supported by the National Key Research and Development Program of China (No. 2017YFA0303700),
	the Major Program of National Natural Science Foundation of China (No. 11690030, 11690032),
	the National Natural Science Foundation of China (No. 61771236),
	the Natural Science Foundation of Jiangsu Province (BK20190297)\\

\bibliographystyle{unsrt}  
\bibliography{references}  

%
%
%
%

\end{document}